\documentclass[preprint]{aastex}
\input psfig.sty

\shortauthors{J.P. Maier, N.M. Lakin, G.A.H. Walker, D.A. Bohlender}
\shorttitle{Interstellar C$_{3}$}

\begin{document}

\title{DETECTION OF C$_{3}$ IN DIFFUSE INTERSTELLAR
CLOUDS}

\author {John P. Maier\altaffilmark{1} and Nicholas M. Lakin}
\affil{Institute for Physical Chemistry, Klingelbergstrasse 80,
University of Basel CH-4053, Switzerland} \email{j.p.maier@unibas.ch,
nicholas.lakin@unibas.ch}

\altaffiltext{1}{Visiting Astronomer, Canada-France-Hawaii Telescope,
operated by the National Research Council of Canada, the Centre
National de la Recherche Scientifique of France, and the University of
Hawaii.}

\author { Gordon A.H. Walker\altaffilmark{1}} \affil{Physics \&
Astronomy Dept., University of British Columbia\\ Vancouver, BC, Canada
V6T 1Z1} \email{walker@astro.ubc.ca}

\author {David A. Bohlender\altaffilmark{1}} \affil{National Research
Council of Canada, Herzberg Institute of Astrophysics\\ 5071 West
Saanich Road Victoria, BC, Canada V9E 2E7}
\email{david.bohlender@nrc.ca}

\begin{abstract} The smallest polyatomic carbon chain, C$_{3}$, has
been identified in interstellar clouds (A$_{v}\sim$1 mag) towards
$\zeta$ Ophiuchi, 20 Aquilae, and $\zeta$ Persei by detection of the
origin band in its $A^{1}\Pi_{u}-X^{1}\Sigma^{+}_{g}$ electronic
transition, near 4052\AA.  Individual rotational lines were resolved up
to $J$=30 enabling the rotational level column densities and
temperature distributions to be determined. The inferred limits for the
total column densities ($\sim$1 to 2$\times10^{12}$ cm$^{-2}$) offer a
strong incentive to laboratory and astrophysical searches for the
longer carbon chains.  Concurrent searches for C$_2^{+}$, C$_2^{-}$ and
C$_3^{-}$ were negative but provide sensitive estimates for their
maximum column densities.  \end{abstract}


\section{Introduction}

Carbon chains play a central role in the chemistry and spectroscopy of
interstellar space.  The detection of cyanopolyacetylenes in dense
interstellar clouds \citep{Kro81} led to the suggestion by
\citet{Dou77} that carbon chain species be considered as candidates for
the diffuse interstellar bands (DIBs), which are found in
the 4000-8500\AA\/ spectral region of stars reddened by interstellar
dust \citep{Her95}.  Since then, many other molecules with a carbon chain 
backbone
have been identified at radio frequencies in dense clouds
\citep{Kaw95}. Meanwhile, advances in laboratory measurements have
provided an understanding of the types and sizes of carbon chains which
have strong electronic transitions in the DIB range \citep{Mai98}.  It
is thus somewhat surprising that as yet among the bare carbon species
only diatomic C$_{2}$ has been identified in
interstellar clouds where DIB are detected.

The 4052\AA\/ electronic band system of C$_{3}$ was first detected in
comets \citep{Swi42} and then in circumstellar shells by infrared
spectroscopy \citep{Hin88}.  Most recently C$_{3}$ was identified
in a dense cloud using sub-mm measurements of its low frequency bending
mode and N$_{\rm tot}$(C$_{3}$)$\sim10^{13}$
cm$^{-2}$ was estimated \citep{Cer00}.  \citet{Sno88} established an upper limit of
5$\times$10$^{10}$ cm$^{-2}$ for the column density of C$_{3}$ in the
direction of $\zeta$ Oph, some two orders of magnitude lower than that
set by \citet{Cle82}.  \citet{Haf95} made a tentative detection of
C$_{3}$ towards an eighth magnitude star in the same part of the sky,
HD 147889, at a column density of 4$\times$10$^{12}$ cm$^{-2}$.
Unfortunately, the star turned out to be a (previously unknown)
double-lined spectroscopic binary which limited their sensitivity.

This letter presents the detection of 
C$_{3}$ towards three stars and infers the column densities in the diffuse
clouds.
Although C$_{2}^{+}$, C$_{2}^{-}$ and C$_{3}^{-}$ were not detected, 
upper limits for their column densities are estimated.
Diatomic species, such as CH,
CN, C$_{2}$, and CH$^{+}$, have been detected towards two of the stars
chosen and their column densities are considered standards with
which to compare models for the physical and chemical processes in
diffuse regions \citep{vanD86}. The four bare carbon chains, C$_{3}$,
C$_{2}^{+}$, C$_{2}^{-}$ and C$_{3}^{-}$, were selected for the present
search because gas phase
electronic transitions for these species have been identified in the laboratory in the 4000-5500\AA\/ region
and their oscillator strengths are known (see table \ref{results}).

\section{The Observations}

Observations of the reddened stars $\zeta$ Oph (HD 149757), $\zeta$ Per
(HD 24398) and 20 Aql (HD 179406) were made with the Gecko echellette
spectrograph on 2000 July 16 and 19, fiber fed from the Cassegrain focus of the
Canada-France-Hawaii 3.6-m telescope (CFHT) \citep{Bau00}.  All three
stars have a visual extinction, A$_{v}$, near 1 and were chosen because
they are bright with sharp interstellar K I lines indicating either
single clouds or little Doppler distortion (in the case of $\zeta$ Oph,
\citet{Cra97} resolved the C$_{2}$ at 8756 \AA\/ into two close
velocity components separated by 1.1 km s$^{-1}$).

The detector was a rear illuminated EEV1 CCD (13.5 $\mu$m$^{2}$
pixels) and the spectral regions were centered at 4047 \AA\/ in the
14th order, and at 5060 \AA\/ and 5400 \AA\/ in the 11th and 10th
orders, respectively.  The ultraviolet Gecko prism was used to
isolate the 14th order, the blue grism for the 11th order, 
while the stock CFHT filter \#1515 was used for the 10th
order observations.  Individual spectra had exposure times ranging from
5 to 20 minutes and were obtained with continuous fiber agitation to
overcome modal noise.  The resulting combined spectra for the
individual stars at each wavelength had unusually high signal-to-noise
ratios (S/N$\sim$800-4000) for CCD observations.  The Th/Ar comparison
arc spectra, taken before and after each spectrograph wavelength
reconfiguration, had a typical FWHM of 2.8 pixels, which corresponds to
resolutions of $R$ = 121000, 113000 and 101000 at
4047, 5060 and 5400 \AA, respectively.

Processing of the spectra was conventional. Groups of biases were taken
several times throughout each night and at each grating setting a
series of flat-field spectra of a quartz-iodide lamp were recorded.
The biases and appropriate flats were averaged and used to remove the
zero-level offset and pixel-to-pixel sensitivity variations of the
CCD.  One-dimensional spectra were extracted using standard IRAF
routines.  Spectra of Vega and other unreddened stars were used to
search for contaminating telluric water vapor lines and stellar
photospheric features.  Heliocentric corrections were applied to each
spectrum.

The observations are summarised in Table \ref{observations} which lists
exposure times and S/N per pixel for each spectral region. The final
column gives the radial velocities measured from the interstellar K I
4044.1 and 4047.2 \AA\/ lines. These velocities have been applied to
each spectrum to put the interstellar features on a laboratory scale.

\section{The Results}

\begin{figure}
\epsscale{0.5}
\centerline{\psfig{figure=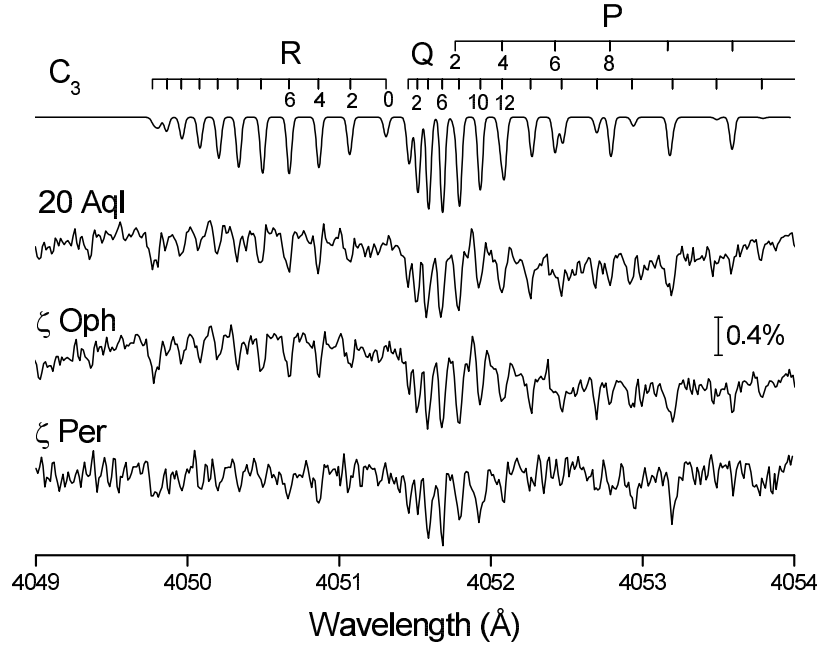,height=8.5cm,angle=-90}}
\caption{
Rotational lines in the origin band of the C$_{3}$
$A^{1}\Pi_{u}-X^{1}\Sigma^{+}_{g}$ transition towards $\zeta$ Oph,
$\zeta$ Per and 20 Aql.   The bar
indicates 0.4\% of the continuum. A known, weak stellar absorption line
at 4053.2 \AA\/ has been removed from the spectrum of $\zeta$ Per.  The
simulated spectrum is for T$_{rot}$= 80 K at a spectral resolution of
110,000.
\label{figure1} }
\end{figure}

The $A^{1}\Pi_{u}-X^{1}\Sigma^{+}_{g}$ origin band of C$_{3}$ is quite
clearly detected towards all three stars.  Figure \ref{figure1} compares the
observed spectra with a simulated C$_{3}$ absorption spectrum
based on the spectrograph resolution and assuming a Boltzmann
distribution in the ground state rotational levels with a temperature
of 80 K. The continuum noise level in the observations is $\sim$0.1 \%.
Low order polynomials have been applied to the stellar data to give a
level continuum (base line) and, in the case of $\zeta$ Per, a weak, broad
stellar feature at 4053.2\AA\/ has been removed. 
Residual broad features in the final spectra are only a few tenths of
a percent deep, much less than in the original, and they in no way mask
the sharp C$_{3}$ lines.
In the simulation the
rotational line intensities were calculated using the H\"{o}nl-London
factors, while the line positions were taken from the laboratory
measurements (this avoids the problem of a perturbation affecting the
low $J$ ground state levels which is not accounted for by the fitted
spectroscopic constants). The individual rotational P, Q and R lines
are clearly resolved in the spectra of all three stars. Table 
\ref{lines} lists the observed 
positions and equivalent widths of each rotational line assigned in 
the spectra of the three stars. The table also gives the corresponding 
positions measured for these transitions in the laboratory 
(Gausset et al. 1965). The positions of some 30 lines in the spectrum 
of $\zeta$ Oph agree 
with the laboratory data to within 0.1 cm$^{-1}$ providing an 
unambiguous identification of C$_{3}$
in these diffuse clouds.

\begin{figure} 
\centerline{\psfig{figure=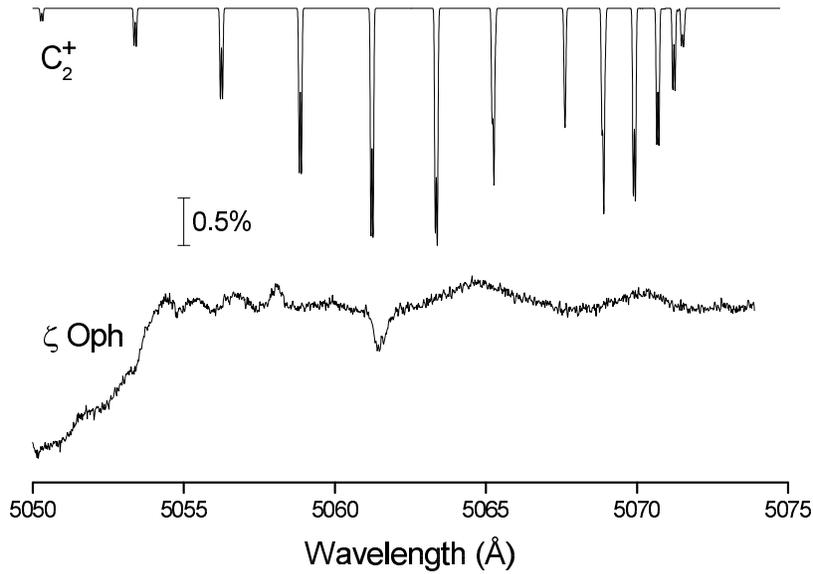,height=7.5cm,angle=-90}}
\caption{A spectrum of $\zeta$
Oph in the region of the C$_{2}^{+}$ $B^{4}\Sigma_{u}^{-}-X^{4}\Sigma^{-}_{g}$
origin band. The upper curve is a
simulated spectrum of C$_{2}^{+}$ showing pairs of
rotational lines. The feature at 5062 \AA\/ cannot be stellar as it
is too narrow but might be a sharp DIB. Note the very high signal-to-noise
($>$ 4000) in the stellar spectrum.  \label{figureC2p} }
\end{figure}

\begin{figure} 
\centerline{\psfig{figure=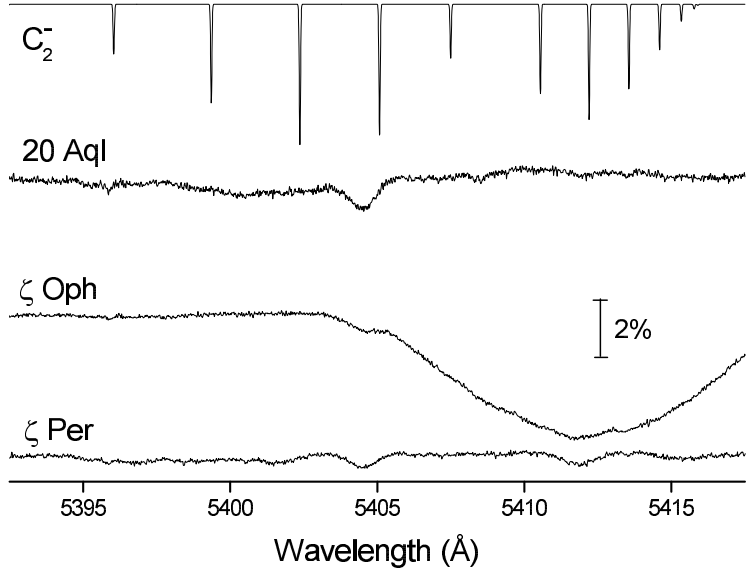,height=7.5cm,angle=-90}}
\caption{ Spectra of $\zeta$
Oph, $\zeta$ Per and 20 Aql in the region of the C$_{2}^{-}$ origin
$B^{2}\Sigma_{u}^{+}-X^{2}\Sigma^{+}_{g}$ band. The upper curve is a simulated 
spectrum of C$_{2}^{-}$ showing
individual rotational lines. The feature at 5404.6 \AA\/ is a known,
diffuse interstellar band (DIB). There is a strong, rotationally
broadened stellar line at 5412 \AA\/ in $\zeta$ Oph.  \label{figureC2m}
} 
\end{figure}

\begin{figure}
\centerline{\psfig{figure=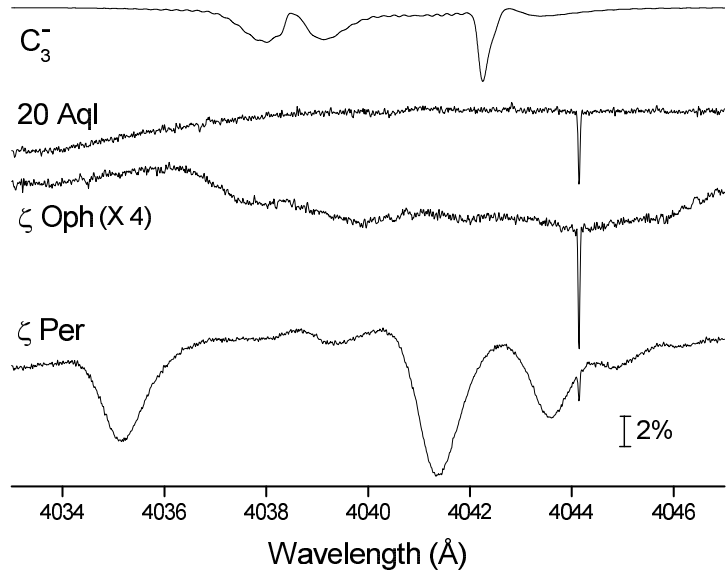,height=7.5cm,angle=-90}}
\caption{ Spectra of $\zeta$ Oph, $\zeta$ Per
and 20 Aql in the region of the $A^{2}\Delta_{u}-X^{2}\Pi_{g}$
C$_{3}^{-}$ origin band. The upper
curve is a simulated spectrum of C$_{3}^{-}$ showing
two unresolved spin-orbit component bands. The spectrum of 
$\zeta$ Per is strongly
contaminated by stellar absorption features.  
The sharp line at 4044.1 \AA\/ is the interstellar K I line which 
was one of two features used to place the spectra on a laboratory 
wavelength. Note that the intensity scale for $\zeta$ Oph is four times
greater than for the other two stars. \label{figureC3m} }
\end{figure}

Figures \ref{figureC2p}, \ref{figureC2m}, and \ref{figureC3m} show the
results of equivalent searches for
$B^{4}\Sigma^{-}_{u}-X^{4}\Sigma^{-}_{g}$ origin band of C$_{2}^{+}$ at
5078.1\AA\/ ($\zeta$ Oph only), the
$B^{2}\Sigma^{+}_{u}-X^{2}\Sigma^{+}_{g}$ origin band of C$_{2}^{-}$ at
5415.9\AA\/ and the $A^{2}\Delta_{u}-X^{2}\Pi_{g}$ origin band of
C$_{3}^{-}$ at 4040.4\AA, together with simulated spectra
for these transitions at 80 K based upon the published spectroscopic
constants 
(\citet{Mai88}, \citet{Her68}, \citet{Tul00}).  In the cases
of C$_2^-$ and C$_2^+$ the linewidth was assumed to be determined by
the spectrograph resolution.
For C$_3^-$ the excited state has been identified as a
short-lived Feshbach resonance and the measured natural linewidth of
$\sim$1 cm$^{-1}$ is employed in the simulation. Weak telluric lines
were been removed from the C$_{2}^{+}$ and C$_{2}^{-}$ observations
using standard procedures.

Unlike the 4050\AA\/ region for C$_{3}$, each of these spectral regions
is contaminated by weak stellar features. Nonetheless, for the
C$_{2}^{+}$ and C$_{2}^{-}$ ions, there are no sharp features
corresponding to the rotational lines in the simulations. On the other
hand, for C$_3^-$ there are features in the spectrum of $\zeta$ Per and
(much less convincing) in the magnified plot for $\zeta$ Oph which
appear to coincide in position and shape with the simulated band
heads.  It is unlikely that these are due to C$_3^-$ because the stellar
lines in $\zeta$ Per have exactly the same shape as the coincident
features. For $\zeta$ Oph the whole C$_3^-$ spectrum sits within a weak
stellar feature (the lines are broadened by rapid rotation $\sim$400
km s$^{-1}$). The photospheric lines in $\zeta$ Oph show nonradial
pulsation `ripples' which will be 'washed out' to some extent by the
long exposure time employed. The spectrum of 20 Aql, which normally 
has the strongest
interstellar lines of the three, is free of stellar features but has no
features coincident with the C$_3^-$ simulation.  It is concluded that,
while interstellar C$_3^-$ might be absent for 20 Aql and present for
the other two stars, it is more likely that in the latter cases the
features are instead stellar.

Table \ref{results} gives the measured equivalent widths for the most
intense C$_{3}$ line (Q(8)) for each star together with an 1$\sigma$
error estimate. For C$_{2}^{+}$, C$_{2}^{-}$ and C$_{3}^{-}$, 3$\sigma$
detection limits are given. The 1$\sigma$ level errors and detection
limits are derived from:

\[W_{\lambda} = (w d)^{\frac{1}{2}} (S/N)^{-1}\]

\noindent where the 1$\sigma$ limiting equivalent width, $W_{\lambda}$,
and the FWHM of the feature, $w$, are both measured in \AA, the
spectrograph dispersion, $d$, in \AA\/ pixel$^{-1}$, and S/N is the
signal to noise per pixel. From the simulations, $w$ = 0.045, 1.0, 0.13
and 0.045\AA\/ for C$_{3}$, C$_{3}^{-}$, C$_{2}^{+}$ and
C$_{2}^{-}$, respectively.

\begin{figure}
\centerline{\psfig{figure=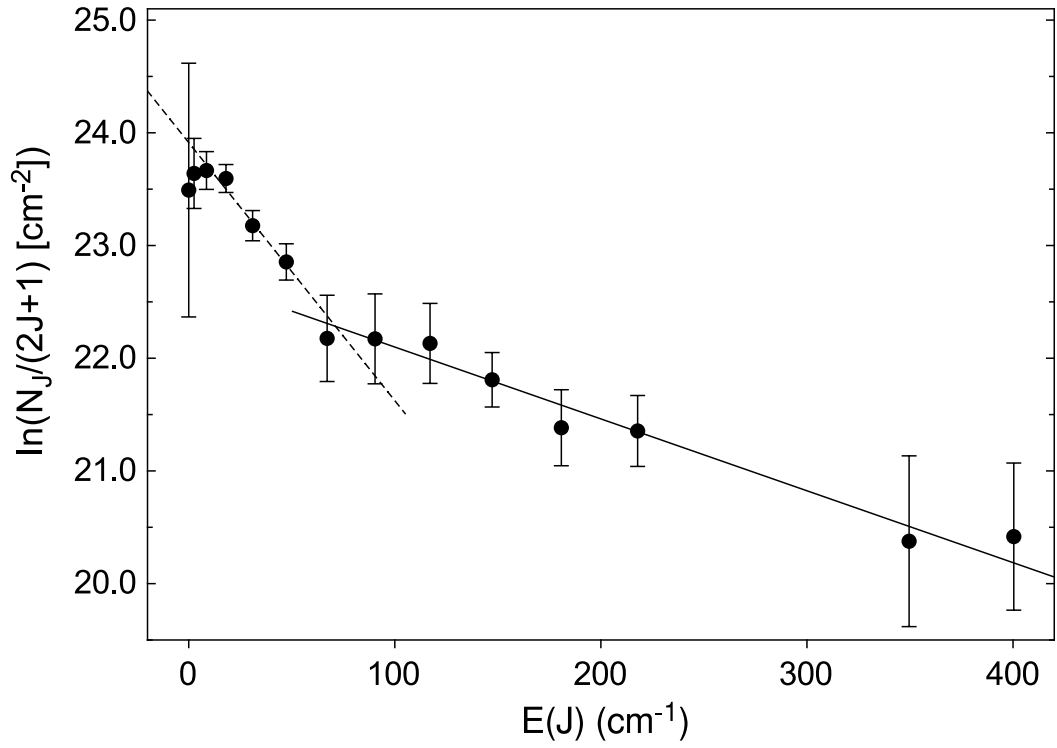,height=10cm,angle=90}}
\caption{
Plot of ln($N_{J}$/(2J+1)) vs. rotational level energy, E(J), of
C$_{3}$ for $\zeta$ Oph.  The error bars are $\pm1 \sigma$ in the measured
column densities.  Straight line fits are shown for J$<$14 (dashed, 60
K) and J$>$14 (solid, 230 K).
\label{figure5} }
\end{figure}

In the case of C$_{3}$, equivalent widths, $W_{\lambda}$, were
determined for each rotational line (varying between
0.3-2.7$\times10^{-4}$\AA) and, in combination with the transition
oscillator strength, f$_{0-0}$, and H\"{o}nl-London factors, the column
densities, $N_{J}$(C$_{3}$), of each rotational level ($J$) in the
ground electronic state were calculated \citep{Lar83}. In cases where
several rotational lines originating from the same level were assigned
(e.g. P(8), Q(8), R(8)) the mean of the determined column densities
was taken.  Figure \ref{figure5} shows a Boltzmann plot of
ln($N_{J}/(2J+1)$) vs. the rotational energy \citep{Gau65} where the
slope is inversely proportional to the rotational temperature.  Among
the lowest rotational levels ($J<$14) the populations are reasonably
approximated by a distribution at 50-70 K, whereas the higher
rotational levels correspond to a temperature of 200-300 K. The
simulation in Figure 1 uses 80 K as this represents an average
temperature for the entire rotational population and allows both the
high and low $J$ lines to be identified.  The high temperature
component of the distribution is apparent in the astronomical spectra
where the R band head and the higher Q lines are more intense than in
the simulation (Figure 1).

\citet{Lut83} also found a bimodal population distribution for C$_{2}$
in diffuse clouds, with similar characteristic temperatures for the low
and high $J$ values. The lower temperature is interpreted as the
kinetic energy of the cloud and for both C$_{2}$ and C$_{3}$ the values obtained
are
comparable to those used in models of diffuse clouds
\citep{vanD86}. The higher temperature component is attributed to
repopulation of the levels in the ground electronic state by radiative
pumping from excited states.  In the case of C$_{3}$ it is expected
that both the $^{1}\Pi_{u}$ state and the higher lying
$^{1}\Sigma_{u}^{+}$ state will contribute to the radiative pumping.

The sensitivity of these measurements is such that $N_{J}$(C$_{3}$) in
the range 0.2-2$\times10^{11}$ cm$^{-2}$ is determined for
rotational levels up to $J$=30.  The $N_{J}$ values were summed to give
the estimated lower limits in the range 1-2$\times10^{12}$ cm$^{-2}$
for the total column density, $N_{\rm tot}$(C$_{3}$), in Table
\ref{results}.

\subsection{Discussion}

A previous search for C$_{3}$ in the direction of $\zeta$ Oph did not 
identify the molecule \citep{Sno88}. It is unclear why this was the case
as, in the light of the present observations, the signal-to-noise quoted for 
these 
measurements was adequate and the upper limit given was some thirty times
lower than the column density reported here.
The present measurements for the column densities of C$_{3}$ are of
the same order of magnitude as the tentative estimate for a translucent 
cloud \citep{Haf95} and can be compared to those of other
polyatomic molecules observed in diffuse interstellar clouds. Column densities
(also towards $\zeta$ Oph) in the 10$^{12-13}$ cm$^{-2}$ range have
been inferred for HCO$^{+}$, C$_{2}$H and C$_{3}$H$_{2}$ from
observations in the mm region by \citet{Luc00}, while H$_{3}^{+}$ has
been identified in diffuse regions and N$_{\rm tot}$(H$_{3}^{+}$)
estimated $\sim4\times10^{14}$ cm$^{-2}$ by \citet{McC98}. Column
densities of C$_{2}$ towards $\zeta$ Oph and $\zeta$ Per have been
determined in the 2-3$\times10^{13}$ cm$^{-2}$ range by \citet{Fed89}.
A current model of the diffuse clouds by \citet{Ruf00} predicts N$_{\rm
tot}$(C$_{2}$)/N$_{\rm tot}$(C$_{3}$) $\sim$ 20 (on a 10$^{5}$ year
time scale), implying N$_{\rm tot}$(C$_{3}$)$\sim10^{12}$
cm$^{-2}$, in agreement with the values deduced from the 
Table \ref{results}. The main production route to C$_{3}$ is presumed
to be the dissociative recombination process:  C$_{3}$H$^{+}$ + $e$
$\rightarrow$ C$_{3}$ + H, where C$_{3}$H$^{+}$ is produced
from smaller species by C$^{+}$ ion insertion.  Under conditions where
ultra-violet radiation penetrates, photodissociation of C$_{3}$ takes
place at a threshold of 1653\AA. As the strong
$^{1}\Sigma_{u}^{+}$$-$$^{1}\Sigma_{g}^{+}$ electronic transition of
C$_{3}$ is predicted to occur around 1700\AA\/ \citep{For96}, the
dissociation process, C$_{3}$ $\rightarrow$ C$_{2}$ + C, may be an
important destruction pathway in diffuse clouds. 

Although only upper limits for the column
densities of C$_{3}^{-}$, C$_{2}^{+}$ and C$_{2}^{-}$ could be presently
established, these species
are of interest as small ionic carbon fragments play a crucial role in
the ion~molecule schemes for diffuse cloud chemistry \citep{Dal76}.
The C$_{2}^{+}$ ion is the only bare carbon cation for which the gas
phase electronic spectrum is known \citep{Mai88}. In diffuse clouds it
is supposed to be the product of the fundamental step: C$^{+}$ + CH
$\rightarrow$ C$_{2}^{+}$ + H. Its main destruction mechanism is
hydrogenation:  C$_{2}^{+}$ + H$_{2}$ $\rightarrow$ C$_{2}$H$^{+}$ + H,
which dominates over recombination with electrons in diffuse regions.
The diffuse cloud model \citep{Ruf00} predicts a C$_{2}^{+}$
abundance  a factor of 10$^{-3}$ lower than C$_{2}$, implying a column
density $\sim10^{10}$ cm$^{-2}$, in accord with the upper limit in
Table \ref{results}.

Published models for diffuse regions do not include the smallest
pure carbon anions, C$_{2}^{-}$ and C$_{3}^{-}$ in their reaction
libraries. Unlike C$_{2}^{+}$, C$_{2}^{-}$ does not react with H$_{2}$
so its main destruction mechanism is expected to be photodetachment.
The similar rotational line widths and oscillator strengths of the
C$_{2}^{-}$ and C$_{2}^{+}$ transitions lead to similar upper limits
for their total column densities.  The width of the unresolved bands for
C$_{3}^{-}$ and the presence of weak stellar features in this spectral
region means that a higher column density of this ion could have
escaped detection. 

The detection of C$_{3}$ provides a powerful incentive for the
laboratory study of the electronic transitions of longer carbon chains
in the gas phase with the aim of comparison with DIB data. The question
as to what types and sizes of carbon chains will have strong
transitions in the 4000-9000\AA\/ range has already been answered: for example,
$^{1}\Sigma_{u}^{+}$$-$$^{1}\Sigma_{g}^{+}$ transitions of the 
odd-number bare chains,
C$_{2n+1}$, $n$=8-30 \citep{Mai98}.  The existence of linear carbon
chains up to C$_{21}$ has been confirmed by the observation of their 
electronic spectra in neon
matrices \citep{Wys99}. As the oscillator strength scales almost
linearly with the length of the molecule, one can expect
f$_{0-0}\sim$10-20 for these carbon chains.  With such an oscillator
strength, a species with a column density $\sim10^{11}$ cm$^{-2}$
would be enough to give rise to a strong DIB, with an equivalent width
of 1\AA.  In view of the column density 
$\sim$10$^{12}$ cm$^{-2}$ for 
C$_{3}$ determined in this work for three diffuse clouds, this
appears to be a reasonable expectation.

\acknowledgements

The support of the Swiss National Science Foundation (project no. 
20-055285.98), the
Canadian Natural Sciences and Engineering Research Council 
and the National Research Council of Canada is gratefully
acknowledged.  The authors thank the staff of the CFHT for their care
in setting up the fiber feed and agitator, thereby making such high
signal-to-noise spectra possible.

\clearpage

\begin{deluxetable}{clccccccccl}
\tabletypesize{\footnotesize}
\tablecaption{The Observations \label{observations} }
\tablewidth{0pt}
\tablehead{
\colhead{star} & \colhead{Sp/L} & \colhead{V} & \colhead{$E_{(B-V)}$} & 
\multicolumn{2}{c}{4045 \AA} &
\multicolumn{2}{c}{5060 \AA} &
\multicolumn{2}{c}{5400 \AA} &
\colhead{K I RV}\\
& & & & 
\colhead{T\tablenotemark{a}} & \colhead{S/N\tablenotemark{b}} &
\colhead{T} & \colhead{S/N} &
\colhead{T} & \colhead{S/N} &
\colhead{(km s$^{-1}$)}
}
\startdata
$\zeta$ Per & B1 Ib   & 2.85 & 0.28 & 4800 & 1200  &   &  &  2700 & 1900 & +13.91 $\pm$0.24  \\
$\zeta$ Oph& O9.5 V  & 2.56 & 0.30 & 5400 & 2400 & 5400 & 4000 &  3000 & 2200 & $-$14.53 $\pm$0.18 \\ 
20 Aql& B3 V    & 5.36 & 0.27 & 10800 & 800 &  &  & 8400 & 900 & $-$12.53 $\pm$0.08 \\
 \enddata
\tablenotetext{a}{Exposure times in seconds.}
\tablenotetext{b}{per pixel.}
\end{deluxetable}

\begin{deluxetable}{cc|lc|lc|lc}
\tabletypesize{\footnotesize}
\tablecaption{The C$_{3}$ Rotational Lines  \label{lines} }
\tablewidth{0pt}
\tablehead{
\colhead{$\lambda_{lab}^{a}$ \AA } &
\colhead{assignment} &
\multicolumn{2}{c}{$\zeta$ Oph} &
\multicolumn{2}{c}{$\zeta$  Per} &	
\multicolumn{2}{c}{20 Aql}\\
\colhead{} &
\colhead{} &
\colhead{$\lambda_{obs}$ \AA} &
\colhead{W$_{\lambda}$ 10$^{-4}$ \AA} &
\colhead{$\lambda_{obs}$ \AA} &
\colhead{W$_{\lambda}$ 10$^{-4}$ \AA} &
\colhead{$\lambda_{obs}$ \AA} &
\colhead{W$_{\lambda}$ 10$^{-4}$ \AA}
}
\startdata
 4049.784& R(22) & &&4049.770$^{b}$&1.016 & 4049.795$^{b}$&1.708\\
 \phantom{40}49.770& R(24) & &&\phantom{40}49.770$^{b}$&1.016 & \phantom{40}49.795$^{b}$&1.708\\ 
 \phantom{40}49.810& R(20)	  & 4049.782$^{b}$	&1.658	  & \phantom{40}49.807$^{b}$&1.162&&\\
 \phantom{40}49.784& R(26)   & \phantom{40}49.782$^{b}$        &1.658    & \phantom{40}49.807$^{b}$&1.162&&\\
   \phantom{40}49.861	& R(18)  & \phantom{40}49.865	&0.309	  & \phantom{40}49.877	&0.511	  
& \phantom{40}49.865	&0.821\\
   \phantom{40}49.963	& R(16)	  & \phantom{40}49.959	&0.726	  & \phantom{40}49.961	&0.773&&\\
   \phantom{40}50.081	& R(14)	  & \phantom{40}50.079	&0.792	  & \phantom{40}50.091	&0.759&&\\
   \phantom{40}50.206	& R(12)	  & \phantom{40}50.198	&0.920	  & \phantom{40}50.198	&0.680	  
& \phantom{40}50.211	&1.165\\
   \phantom{40}50.337	& R(10)	  & \phantom{40}50.329	&1.034	  & \phantom{40}50.342	&0.679	  & \phantom{40}50.339	
&1.450\\
   \phantom{40}50.495	& R(8)	  & \phantom{40}50.483	&1.525	  & \phantom{40}50.497	&0.383	  & \phantom{40}50.489	
&2.204\\
   \phantom{40}50.670	& R(6)	  & \phantom{40}50.669	&1.562	  & \phantom{40}50.662	&0.943	  & \phantom{40}50.667	
&2.330\\
   \phantom{40}50.865	& R(4)	  & \phantom{40}50.863	&1.018	  & \phantom{40}50.864	&1.068	  & \phantom{40}50.853	
&2.225\\
   \phantom{40}51.069	& R(2)	  & \phantom{40}51.073	&0.896   &&                       & \phantom{40}51.074	&1.657\\
   \phantom{40}51.309	& R(0)	  & \phantom{40}51.267$^{c}$ &0.371	 &&   & \phantom{40}51.386$^{c}$&0.730\\
   \phantom{40}51.461	& Q(2)	  & \phantom{40}51.457	&1.045	  & \phantom{40}51.457	&0.923	  & \phantom{40}51.455	
&1.180\\
   \phantom{40}51.521	& Q(4)	  & \phantom{40}51.515	&2.187	  & \phantom{40}51.518	&0.752	  & \phantom{40}51.506	
&1.876\\
   \phantom{40}51.590	& Q(6)	  & \phantom{40}51.586	&2.719	  & \phantom{40}51.588	&2.183	  & \phantom{40}51.583	
&1.965\\
   \phantom{40}51.682	& Q(8)	  & \phantom{40}51.679	&2.336	  & \phantom{40}51.680	&2.016	  & \phantom{40}51.669	
&2.229\\
   \phantom{40}51.793	& Q(10)	  & \phantom{40}51.788	&2.138	  & \phantom{40}51.795	&2.294	  & \phantom{40}51.787	
&2.508\\
   \phantom{40}51.929	& Q(12)	  & \phantom{40}51.930	&1.060	  & \phantom{40}51.922	&2.020	  & \phantom{40}51.929	
&2.186\\
   \phantom{40}52.062& P(4)& \phantom{40}52.074$^{b}$&1.266& \phantom{40}52.085$^{b}$&1.043& \phantom{40}52.065$^{b}$
&3.130\\
   \phantom{40}52.089& Q(14)& \phantom{40}52.074$^{b}$&1.266& \phantom{40}52.085$^{b}$&1.043& \phantom{40}52.065$^{b}$
&3.130\\ 
   \phantom{40}52.271	& Q(16)	  & \phantom{40}52.262	&1.217			  &&& \phantom{40}52.271	&1.821\\
   \phantom{40}52.424	& P(6)	&&				&&  & \phantom{40}52.456	&2.499\\
   \phantom{40}52.473	& Q(18)	  & \phantom{40}52.466	&1.233&&&&\\
   \phantom{40}52.698	& Q(20)	  & \phantom{40}52.701	&0.923&&&&\\
   \phantom{40}52.792	& P(8)	  & \phantom{40}52.784	&0.605		&&	  & \phantom{40}52.772	&1.592\\
   \phantom{40}52.900  	& Q(22)	  & \phantom{40}52.929	&0.985		&&	  & \phantom{40}52.939	&1.122\\
   \phantom{40}53.180	& P(10)	  & \phantom{40}53.197$^{b}$&1.883&&&\\
   \phantom{40}53.207       & Q(24)   & \phantom{40}53.197$^{b}$&1.883&&&\\
   \phantom{40}53.590	& P(12)	  & \phantom{40}53.593	&0.678		&&	  & \phantom{40}53.588	&1.237\\
   \phantom{40}53.795	& Q(28)	  & \phantom{40}53.786	&0.469		&&	  & \phantom{40}53.781	&1.098\\
   \phantom{40}54.112	& Q(30)	  & \phantom{40}54.113	&0.523&&&&\\
   \phantom{40}54.459	& P(16)	  & \phantom{40}54.445	&0.870&&&&\\
   \phantom{40}54.908	& P(18)	  & \phantom{40}54.904	&1.122&&&&\\
 \enddata
\tablenotetext{a}{\citet{Gau65}}
\tablenotetext{b}{These features are assigned to two blended lines of
similar intensity in the simulation and are not used to determine
rotational level column densities.}
\tablenotetext{c}{ R(0) is a very weak line with a correspondingly
larger positional error compared to stronger neighboring lines.}
\end{deluxetable}

\begin{deluxetable}{clcccccl} \tabletypesize{\footnotesize}
\tablecaption{Total and rotational level column densities of C$_{3}$
towards the three stars and estimated upper limits for the equivalent
widths and column densities of C$_{3}^{-}$, C$_{2}^{+}$ and C$_{2}^{-}$
\label{results}
 }

\tablewidth{0pt} 
\tablehead{ \colhead{species} & 
\colhead{transition} &
\colhead{$\lambda_{0-0}$} & 
\colhead{f$_{0-0}$} & 
\colhead{star} &
\colhead{$W_{\lambda}$\tablenotemark{a}} &
\colhead{$N_{J}$\tablenotemark{a}} & 
\colhead{$N_{\rm tot}$}\\ & &
\colhead{(\AA)} & & & 
\colhead{10$^{-4}$\AA\/} & 
\colhead{10$^{10}$
cm$^{-2}$} & 
\colhead{10$^{12}$ cm$^{-2}$} 
} 
\startdata
C$_{3}$&$A^{1}\Pi_{u}-X^{1}\Sigma^{+}_{g}$&4051.5\tablenotemark{b}&0.016\tablenotemark{c}&$\zeta$ Oph&2.3$\pm$0.1&20&1.6\\ 
&&&&20 Aql&2.2$\pm$0.3&19&2.0\\ 
&&&&$\zeta$ Per& 2.0$\pm$0.2&17&1.0\\\hline
C$_{3}^{-}$&$A^{2}\Delta_{u}-X^{2}\Pi_{g}$&4040.4\tablenotemark{d}&0.04\tablenotemark{e}&$\zeta$ Oph&6.0&&$<$0.3\\ 
&&&&20 Aql&20&&$<$1.2\\ 
&&&&$\zeta$ Per&12&&$<$0.7\\\hline
C$_{2}^{+}$&$B^{4}\Sigma^{-}_{u}-X^{4}\Sigma^{-}_{g}$&5066.9\tablenotemark{f}&0.025\tablenotemark{f}&$\zeta$ Oph&0.35&$<$1.1&$<$0.04\\\hline
C$_{2}^{-}$&$B^{2}\Sigma^{+}_{u}-X^{2}\Sigma^{+}_{g}$&5408.6\tablenotemark{g}&0.044\tablenotemark{h}&$\zeta$ Oph& 0.30&$<$0.5&$<$0.02\\ 
&&&&20 Aql&0.55&$<$0.9&$<$0.03\\ 
&&&&$\zeta$ Per&0.35&$<$0.6&$<$0.02\\
 \enddata \tablenotetext{a}{For C$_{3}$ the $N_{J}$ values refer to the
J=8 level and $W_{\lambda}$ to the most intense Q(8) line. For
C$_{2}^{-}$ and C$_{2}^{+}$ $W_{\lambda}$ are 3$\sigma$ detection
limits, based on simulated spectra at the spectrograph resolving power,
and the $N_{J}$ values are determined from the most intense R lines at
80 K. For C$_{3}^{-}$ individual rotational lines are not resolved and
$W_{\lambda}$ is the 3$\sigma$ detection limit for the blended R
branch.} \tablenotetext{b}{\citet{Gau65}}
\tablenotetext{c}{\citet{Bec79}} \tablenotetext{d}{\citet{Tul00}}
\tablenotetext{e}{Estimated value
\citep{Tul00}}
\tablenotetext{f}{\citet{Mai88}} \tablenotetext{g}{\citet{Her68}}
\tablenotetext{h}{\citet{Leu82}} \end{deluxetable}

\clearpage

\end{document}